\documentclass[aps,prd,groupedaddress,tighten]{revtex4}
\usepackage{amssymb}
\tighten

\usepackage{graphicx}
\usepackage{bm}
\usepackage{amsmath}

\newcommand{\rp}{r_{+}}

\input{tcilatex}

\begin{document}

\title{Regular black holes in quadratic gravity}
\author{Waldemar Berej}
\affiliation{Institute of Physics, 
Maria Curie-Sk\l odowska University\\
pl. Marii Curie-Sk\l odowskiej 1, 
20-031 Lublin, Poland}
\author{Jerzy Matyjasek}
\email{jurek@kft.umcs.lublin.pl, matyjase@tytan.umcs.lublin.pl}                                         
\affiliation{Institute of Physics, 
Maria Curie-Sk\l odowska University\\
pl. Marii Curie-Sk\l odowskiej 1, 
20-031 Lublin, Poland}
\author{Dariusz Tryniecki}
\affiliation{Institute of Theoretical Physics, 
Wroc\l aw University\\
pl. M. Borna 9, 
50-204 Wroc\l aw, Poland}
\author{Mariusz Woronowicz} 
\affiliation{Institute of Theoretical Physics, 
Wroc\l aw University\\
pl. M. Borna 9, 
50-204 Wroc\l aw, Poland}

\date{\today}

\begin{abstract}
The first-order correction of the perturbative solution of the coupled
equations of the quadratic gravity and nonlinear electrodynamics is
constructed, with the zeroth-order solution coinciding with the ones
given by Ay\'on-Beato and Garc\'\i a and by Bronnikov. It is shown that
a simple generalization of the Bronnikov's electromagnetic Lagrangian
leads to the solution expressible in terms of the polylogarithm
functions. The solution is  parametrized by two integration constants
and depends on two free parameters. By the boundary conditions the
integration constants are related to the charge and total mass of the
system as seen by a distant observer, whereas the free parameters are
adjusted to make the resultant line element regular at the center. It
is argued that various curvature invariants are also regular there
that strongly suggests the regularity of the spacetime. Despite the
complexity of the problem the obtained solution can be studied
analytically. The location of the event horizon of the black hole, its
asymptotics and temperature are calculated. Special emphasis is put on
the extremal configuration.
\end{abstract}


\maketitle

%\preprint{}

%Title of paper
%%%%%%%%%%%%%%%%%%%%%%%%%%%%%%%%%%%%%%%%%%%%%%%
%%%%%%%%%%%%%%%%%%%%%%%%%%%%%%%%%%%%%%%%%%%%%%%

%%%%%%%%%%%%%%%%%%%%%%%%%%%%%%%%%%%%%%%%%%%%%%%
%%%%%%%%%%%%%%%%%%%%%%%%%%%%%%%%%%%%%%%%%%%%%%%

%%%%%%%%%%%%%%%%%%%%%%%%%%%%%%%%%%%%%%%%%%%%%%%

%%%%%%%%%%%%%%%%%%%%%%%%%%%%%%%%%%%%%%%%%%%%%%%
%%%%%%%%%%%%%%%%%%%%%%%%%%%%%%%%%%%%%%%%%%%%%%%
%%%%%%%%%%%%%%%%%%%%%%%%%%%%%%%%%%%%%%%%%%%%%%%

% insert suggested PACS numbers in braces on next line

% insert suggested keywords - APS authors don't need to do this

%%%%%%%%%%%%%%%%%%%%%%%%%%%%%%%%%%%%%%%%%%%%%%
%\section{\label{sec:Intro} Introduction}
%%%%%%%%%%%%%%%%%%%%%%%%%%%%%%%%%%%%%%%%%%%%%%

\section{Introduction}

One of the most important and intriguing questions of the black hole
physics is the problem of singularities that reside in their internal
region, hidden to an external observer by the event horizon. In the
vast majority of papers singularities are treated as symptoms of
illness of the theory rather than its health (see however
Ref.~\cite{Horowitz} for a different point of view), and, consequently, a
great deal of efforts were directed to constructing singularity-free
models. However, a subtle point is that the Einstein field equations
loose their predicative power and cannot be trusted when the curvature
of the manifold approaches the Planck regime. Indeed, according to our
present understanding a proper description of the gravitational
phenomena should be given by the quantum gravity, being perhaps a part
of a more fundamental theory. And although at the present stage we
have no clear idea how this theory looks like, we expect that the
action functional describing its low-energy approximation should
consist of higher order terms constructed from the curvature and its
covariant derivatives to some required order. It means that in the
full theory the analogs of the nonsingular solution of the Einstein
gravity may loose their nonsingular character as well as the singular
ones their singularities. 

Among various modifications of the general relativity the prominent
role is played by the quadratic gravity
\cite{Utiyama,Stelle1,Stelle2,Weyl1,Weyl2,Pauli,sirArthur}.
Motivations for introducing into the action functional terms which are
quadratic in curvature are numerous. For example, when invented, the
equations of quadratic gravity have been treated as an exact
formulation of the theory of gravitation. For historic informations
and important references the reader is referred to Ref. \cite{HJS}. It
may be considered, quite naturally,  as truncation of series expansion
of the action of the more general theory. 
Such terms appear generically in one-loop calculations of the quantum
field theory in curved background~\cite{Birrell}.
Moreover, from the point of
view of the semi-classical gravity it might be treated (in certain
circumstances) as some sort of
a poor man's stress- energy tensor, allowing in a relatively simple
way to mimic, especially when the application of the full stress-
energy tensor would produce extremely complicated or even intractable
results, the fairly more complex source term of the field equations.

Analyses of the spherically-symmetric and static solutions to the
higher derivative theory has been carried out in \cite
{Stelle1,Lousto1,Lousto2,Lousto3,Holdom1,Tryn}. For example, in
Ref~\cite {Stelle1} it has been shown that the weak-field limit of the
quadratic gravity involves, beside the ordinary Newtonian term, also
the terms with the Yukawa-like potential. Series solutions near the
$r=0$ have been investigated in \cite{Stelle1,Holdom1}. Such solutions
are limited to the closest neighborhood of the center and should be
matched to the appropriate solutions valid for large $r$ that require
numerical integration. However, an important lesson that follows from
this calculations is that the regular solutions of the equations of
the higher order theories are quite common \cite{Stelle1}.

As compared with the General Relativity, the Lagrangian of the
quadratic gravity in four dimensions requires two additional terms
$\alpha R_{ab}R^{ab}+\beta R^{2},$ where $\alpha $ and $\beta $ are
the coupling constants.  Since the
coupling constants are expected to be very small, one can easily device
a perturbative approach to the problem treating the classical solution
of the Einstein field equations as the zeroth-order of the
approximation. Successive perturbations are therefore solutions of the
differential equations of ascending complexity. It should be noted
that although the method is clear the calculations beyond the first-
order may be intractable.

According to the well-known theorem, if the Lagrangian $\mathcal{L}
\left( F\right) $ has Maxwell asymptotics for weak fields, i.e.,
$\mathcal{L}\left( F\right) \sim F$ and 
$\mathcal{L}_{F}=d\mathcal{L}/dF\rightarrow $ $1$ as $ F\rightarrow 0$, 
then any static and spherically-symmetric solutions to the system of
coupled equations of nonlinear electrodynamics and general relativity
characterized by electric charge, $Q_{e}$, cannot have a regular
center \cite{Bronnikov3,Bronnikov1,Bronnikov2}. This no-go theorem
does not forbid existence of the solutions with magnetic charge, $Q$,
as well as some hybrid solutions in which the electric field does not
extend to the central region. On the other hand, however, it has been
argued in Ref. \cite {Burinskii1} that the Maxwell asymptotics at
great distances are essentially different from these at the center,
and, consequently, the condition $ \mathcal{L}_{F}\rightarrow 1$ as
$F\rightarrow 0$ is too restrictive.

The issue of the regular black holes in general relativity has a long
and interesting history. For example, one of the method that can be used
in construction of such configurations is simply replacing the 
singular black hole interior by a regular core.
This idea appeared almost forty years
ago, in mid sixties~\cite{Sakharov,Gliner,Bardeen:68} and is actively
investigated today. 
In models considered in Refs.~\cite{frolov1,frolov2} part 
of the region inside the event horizon
is joined through a thin boundary layer to de Sitter geometry.
Such a geometric surgery certainly does not exhaust all interesting possibilities:
of equal importance are the regular geometries with suitably 
chosen profile functions, or, better,
exact solutions constructed for specific, physically reasonable 
sources~\cite{Irina1,Irina2,Borde,Mars,Ayon-Beato:2000zs,Ayon-Beato:2004ih}.
One of the most intriguing regular solutions to the coupled equations
of the nonlinear electrodynamics and gravity have been constructed by
Ay\'on-Beato and Garc\'\i a \cite{ABG} and by Bronnikov
\cite{Bronnikov1}. We shall refer to solutions of this type as ABGB
geometry. The former describes a regular static and spherically
symmetric configuration with an electric charge whereas the latter
describes a similar geometry characterized by the total mass $M$ and $
Q$. For certain values of the parameters both solutions describe a
black hole. It should be noted that the electric solution does not
contradict the non existence theorem as the formulation of nonlinear
electrodynamics employed by Ay\'on-Beato and Garc\'\i a ($\mathcal{P}$
framework in the nomenclature of Refs. \cite{Bronnikov1,Bronnikov2})
is not the one to which one refers in the proviso of the no-go
theorem. Indeed, the solution of Ay\'on-Beato and Garc\'\i a has been
constructed in a formulation of the nonlinear electrodynamics obtained
from the original one ($\mathcal{F}$ framework) by means of a
Legendre transformation (see Ref.~\cite{Bronnikov2} for details). An
attractive feature of this solutions that certainly simplifies
calculations is possibility to express the location of the horizons in
terms of the known special functions \cite{Kocio1,Kocio2}.

The natural question that arises in connection with the foregoing
discussion is whether or not it is possible to construct an analog of
the solution of the ABGB-type in the quadratic gravity which
shares with its classical counterpart regularity at $r=0.$ And although
the full, detailed answer is beyond our capabilities, it is possible
to provide an affirmative answer to the restricted problem. Indeed,
since the complexity of the coupled equations of the quadratic gravity
and nonlinear electrodynamics, even in the simplest case of
spherically-symmetric and static geometries, hinders
construction of the exact solution, one has to refer to the analytical
approximations or numerical methods.

In this paper we shall employ perturbative methods to construct the
first-order solution to the equations of the quadratic gravity with
the source term being generalization of the stress-energy tensor of
the Bronnikov type. The zeroth-order solution coincides, as expected,
with the ABGB
line element whereas the first-order correction can be elegantly
expressed in terms of the polylogarithm functions. An interesting
feature of this very solution is its regularity for $r = 0.$
The Kretschmann scalar and other curvature invariants are
also regular at the center that suggests regularity of the underlying 
geometry. It should be emphasized that the
demonstration of the regularity of the full solution would require
either profound understanding of the perturbation series to any
required order or construction of the full, physically acceptable
nonperturbative solution.
On the other hand, the perturbative approach is expected to yield
reasonable results especially for higher derivative dynamical
equations. In fact it may be the only method to deal with
them. Indeed, since the quadratic gravity involves fourth-order
derivatives of the metric their nonperturbative solutions may appear
to be spurious and one has to invent a  method for systematic selecting physical
ones. 
%%%%%%%%%%%%%%%
It seems that the acceptable solutions, when expanded 
in powers of the small parameter,
should reduce to those  obtained within the framework 
of  perturbative approach~\cite{Simon1,Simon:1991jn,Parker1}.
 
%%%%%
The paper is organized as follows. In Sec.~\ref{sec2} we introduce
basic equations  and briefly sketch employed method. We choose the
line element in the form propounded by Visser~\cite{Matt1}, which has
proved to be a very useful representation considerably simplifying calculations.
In Sec.~\ref{sec3} we introduce the Lagrangian of the nonlinear 
electrodynamics, construct solutions to the first-order equations
and establish regularity of the thus obtained line element.
Subsequently we study the limit of the vanishing $Q$ and analyse
the regularity of various curvature invariants. Corrections to the 
location of the inner and outer horizons and to  temperature are
given in Sec.~\ref{sec4}.
The position of the horizons of the ABGB spacetime is given
in terms of the real branches of the Lambert functions.
The extremal configuration is studied in Sec.~\ref{sec5}. Specifically,
we calculate modifications of the location of the degenerate horizon
caused by quadratic gravity and analyse relations between $Q$ and the 
total mass as seen by a distant observer.  
Sec.~\ref{sec6} contains final remarks. In Appendix  we have collected
useful formulas and presented a brief description of the method of integration 
of the field equations in terms of the polylogarithm functions.
Throughout the paper the geometric system of units has been adopted and
the the sign conventions are taken to be that of MTW~\cite{MTW}.

\section{The equations}
 \label{sec2}

In absence of the cosmological term, the coupled system of the nonlinear 
electrodynamics and the quadratic gravity is described by the action 
\begin{equation}
S=\frac{1}{16\pi G}S_{g}+S_{m},
\end{equation}
where 
\begin{equation}
S_{g}=\int \left( R+\alpha R^{2}+\beta R_{ab}R^{ab}\right) \sqrt{-g}\,d^{4}x,
\end{equation}
and 
\begin{equation}
S_{m}=-\dfrac{1}{16\pi }\int \mathcal{L}\left( F\right) \sqrt{%
-g}\,d^{4}x.
\end{equation}
Here $\mathcal{L}\left( F\right) $ is some functional of $F=F_{ab}F^{ab}$
(its exact form will be given later) and all symbols have their usual
meaning. The third possible term constructed form the
Kretschmann scalar, $\gamma R_{abcd}R^{abcd}$, may by removed from the
Lagrangian with the help of the Gauss-Bonnet invariant.
Of numerical parameters $\alpha $ and $\beta $ we assume, as usual, that they
are small and of comparable order,
otherwise they would lead to the observational consequences
within our solar system. 
Their ultimate values should be 
determined form observations of light deflection, binary
pulsars and cosmological data~\cite{Odylio,accioly:2001cc,Mijic:1986iv}. 
Moreover, following Ref.~\cite{Lousto3}, we shall restrict ourselves 
to spacetimes of small curvatures,
for which the conditions
\begin{equation}
|\alpha R| <<1, \hspace{5mm}    |\beta R_{ab}|<<1
\end{equation}
hold. 
Although demanding that the mass scales associated with the linearized
equations are real may place additional 
constraints~\cite{Steve,Whitt:1984pd,Audretsch:1993kp} 
on $\alpha$ and $\beta,$
we shall treat them as small but arbitrary.

The tensor $F_{ab}$ satisfies equations

\begin{equation}
\nabla _{a}\left( \dfrac{d\mathcal{L}\left( F\right) }{dF}F^{ab}\right) =0,
\end{equation}
\begin{equation}
\nabla _{a}\,^{\ast }F^{ab}=0,
\end{equation}
and the asterix denotes the Hodge duality. The stress-energy tensor defined
as 
\begin{equation}
T^{ab}=\frac{2}{\sqrt{-g}}\frac{\delta }{\delta g_{ab}}S_{m}  \label{tensep}
\end{equation}
is given therefore by 
\begin{equation}
T_{a}^{b}=\dfrac{1}{4\pi }\left( \dfrac{d\mathcal{L}\left(
F\right) }{dF}F_{ca}F^{cb}-\dfrac{1}{4}\delta _{a}^{b}\mathcal{L}\left(
F\right) \right) .
\end{equation}
Differentiating functionally the action $S$ with respect to the metric
tensor one has 
\begin{equation}
L^{ab}=G^{ab}-\alpha I^{ab}-\beta J^{ab}=8\pi T^{ab},  \label{2nd_order}
\end{equation}
where 
\begin{equation}
I^{ab}=2R^{;\,ab}-2RR^{ab}+\frac{1}{2}g^{ab}\left( R^{2}-4\Box R\right)
\end{equation}
and 
\begin{equation}
J^{ab}=R^{;\,ab}-\Box R^{ab}-2R_{cd}R^{cbda}+\frac{1}{2}g^{ab}\left(
R_{cd}R^{cd}-\Box R\right) .
\end{equation}

Let us consider the spherically symmetric and static configuration described
by the line element of the form 
\begin{equation}
ds^{2}=-e^{2\psi \left( r\right)}f(r) dt^{2}+\frac{dr^{2}}{f(r) } + r^{2}d\Omega
^{2}  \label{el_gen},
\end{equation}
where
\begin{equation}
f(r) = 1 - \frac{2 M(r)}{r}.
\end{equation}
The spherical symmetry places restrictions on the components of $F_{ab}$
tensor and its only nonvanishing components compatible with the assumed
symmetry are $F_{01}$ and $F_{23}$. Simple calculations yield 
\begin{equation}
F_{23}=Q\sin \theta
\end{equation}
and 
\begin{equation}
r^{2}e^{-2\psi }\dfrac{d\mathcal{L}\left( F\right) }{dF}F_{10}=Q_{e},
\end{equation}
where $Q$ and $Q_{e}$ are the integration constants interpreted as the
magnetic and electric charge, respectively. In the latter we shall assume
that the electric charge vanishes, and, consequently, $F$ is given by 
\begin{equation}
F=\dfrac{2Q^{2}}{r^{4}}.  \label{postacF}
\end{equation}
The stress-energy tensor (\ref{tensep}) calculated \ for this configuration
is 
\begin{equation}
T_{t}^{t}=T_{r}^{r}=-\dfrac{1}{16\pi }\mathcal{L}\left( F\right)  \label{t1}
\end{equation}
and 
\begin{equation}
T_{\theta }^{\theta }=T_{\phi }^{\phi }=\dfrac{1}{4\pi }\dfrac{d\mathcal{L}
\left( F\right) }{dF}\dfrac{Q^{2}}{r^{4}}-\dfrac{1}{16\pi }\mathcal{L}\left(
F\right) .
\end{equation}

To simplify calculations and to keep control of the order of terms in
complicated series expansions we shall introduce a dimensionless parameter $
\varepsilon $ substituting $\alpha \rightarrow \varepsilon \alpha $ and $
\beta \rightarrow \varepsilon \beta $. We shall put $\varepsilon =1$ at the
final stage of calculations. Of functions $M\left( r\right) $ and $\psi
\left( r\right) $ we assume that they can be expanded as 
\begin{equation}
M\left( r\right) =M_{0}\left( r\right) +\varepsilon M_{1}\left( r\right) +
\mathcal{O}\left( \varepsilon ^{2}\right)  \label{Mser}
\end{equation}
and 
\begin{equation}
\psi \left( r\right) =\varepsilon \psi _{1}\left( r\right) +\mathcal{O}
\left( \varepsilon ^{2}\right) .  \label{psiser}
\end{equation}

Consider the left hand side of Eq. (\ref{2nd_order}) calculated for the line
element (\ref{el_gen}) first. Making use of the above expansions and
collecting the terms with the like power one obtains 
\begin{equation}
L_{t}^{t}=-\frac{2}{r^{2}}(M_{0}^{\prime }+\varepsilon M_{1}^{\prime
}-\varepsilon S_{t}^{t}),  \label{1st}
\end{equation}
where 
\begin{align}
S_{t}^{t}& =\beta \left( \frac{2\,M_{0}^{\prime }}{r^{2}}-\frac{
8\,M_{0}\,M_{0}^{\prime }}{r^{3}}+\frac{2\,{M_{0}^{\prime }}^{2}}{r^{2}}-
\frac{2\,M_{0}^{\prime \prime }}{r}+\frac{5\,M_{0}\,M_{0}^{\prime \prime }}{
r^{2}}-\frac{M_{0}^{\prime }\,M_{0}^{\prime \prime }}{r}\right.  \notag \\
& \left. +\frac{{M_{0}^{\prime \prime }}^{2}}{2}+M_{0}^{(3)}-\frac{
M_{0}\,M_{0}^{(3)}}{r}-M_{0}^{\prime
}\,M_{0}^{(3)}+r\,M_{0}^{(4)}-2\,M_{0}\,M_{0}^{(4)}\right)  \notag \\
& -\alpha \left( \frac{24\,M_{0}\,M_{0}^{\prime }}{r^{3}}-\frac{
8\,M_{0}^{\prime }}{r^{2}}-\frac{4\,{M_{0}^{\prime }}^{2}}{r^{2}}+\frac{
8\,M_{0}^{\prime \prime }}{r}-\frac{18\,M_{0}\,M_{0}^{\prime \prime }}{r^{2}}
-{M_{0}^{\prime \prime }}^{2}\right.  \notag \\
& \left. +\frac{2\,M_{0}^{\prime }\,M_{0}^{\prime \prime }}{r}
-4\,M_{0}^{(3)}+\frac{6\,M_{0}\,M_{0}^{(3)}}{r}+2\,M_{0}^{\prime
}\,M_{0}^{(3)}-2\,r\,M_{0}^{(4)}+4\,M_{0}\,M_{0}^{(4)}\right)   \notag \\
&  \label{1sta}
\end{align}
and $M_{0}^{\prime},$ $M_{0}^{\prime \prime}$ and $M^{(i)}_{0}$ for $i \geq 3 $ denote 
first, second and $i-$th derivatives with respect to the radial coordinate. 
On the other hand, a simple combination of the components of $L_{a}^{b}$
tensor 
\begin{equation}
L_{r}^{r}-L_{t}^{t}=0  \label{2nd}
\end{equation}
can be easily integrated to yield 
\begin{equation}
\psi _{1}(r)\,=\,(2\alpha +\beta )M_{0}^{(3)}-{\frac{4}{r^{2}}}(3\alpha
+\beta )M_{0}^{\prime }+C_{1},  \label{pseq}
\end{equation}
where $C_{1}$ is the integration constant. 
It should be noted that contrary to the case of coupled system of the Maxwell 
equations and quadratic gravity considered in 
Refs.~\cite{Lousto1,Lousto2,Lousto3,Tryn}
now we have explicit dependence on the parameter $\alpha.$ This together 
with the nonlinear character of the source term (that will be specified below)
results in substantial complications of the first-order equations.

To develop the model further we
shall express solutions of the system of differential equations consisting
of the time component of Eqs. (\ref{2nd_order}) and Eq. (\ref{2nd}) in terms
of the total mass $\mathcal{M}$ as seen by a distant observer 
\begin{equation}
\lim_{r\rightarrow \infty }M\left( r\right) =\mathcal{M,}  \label{bound1}
\end{equation}
whereas for the function $\psi \left( r\right) $ we shall adopt the natural
condition 
\begin{equation}
\lim_{r\rightarrow \infty }\psi \left( r\right) =0.  
\label{bound2}
\end{equation}

\section{Solutions}
 \label{sec3}

Further considerations require specification of the Lagrangian 
$\mathcal{L}\left( F\right) .$ 
Following Ay\'on-Beato, Garc\'\i a and Bronnikov let us chose it in the
form 
\begin{equation}
\mathcal{L}\left( F\right) \,=F\left[ 1-\tanh ^{2}\left( s\,\sqrt[4]{\frac{
Q^{2}F}{2}}\right) \right] ,  \label{labg}
\end{equation}
where 
\begin{equation}
s=\frac{\left| Q\right| }{2b},  \label{sabg}
\end{equation}
and the free parameter $b$ will be adjusted to guarantee regularity at
the center. 

Before proceeding further we shall briefly discuss the question of
regularity, postponing presentation of the technical details for a while. 
First, observe that the zeroth-order solution coincides with a general 
ABGB line element that depends
on a free parameter $b$ and an integration constant $C_{2}$. 
On the other hand, the first-order solution 
written in the suitable representation 
approaches, as we shall see, a constant value, 
say $\mu ,$ as $r \to \infty,$ and, consequently, by the boundary 
conditions (\ref{bound1}) one has $C_{2} = \mathcal{M} - \varepsilon \mu.$ 
The thus obtained line element is generally singular at the center,  and
the only way  to make it regular consists in appropriate choice
of the free parameter.
The regularity is understood here as the regularity of the
line element rather than regularity of the spacetime itself as the
latter requires the curvature invariants to be regular. We shall
return to this issue later. 

%%%%%%%%%%%%%%%%%%%%%%%%%%%%%%%%%%%%%%%%%%%%%%%%%%%%%%%%%%%%%%%%%%%%%%%%
%%%%%%%%%%%%%%%%%%%%%%%%%%%%%%%%%%%%%%%%%%%%%%%%%%%%%%%%%%%%%%%%%%%%%%%%
Now we present the calculations in a more systematic form. 
In order to guarantee sufficient generality of our considerations 
we shall take the parameter $b$ in
the form
\begin{equation}
b=b_{1}+\varepsilon b_{2},  \label{bgen}
\end{equation}
with $b_{2}\neq 0.$ \textit{\ }
%%%%%%%%%%%%%%%%%%%%%%%
Since we have assumed the expansions of $M\left( r\right) $ and $\psi \left(
r\right) $ in the form given by Eqs. (\ref{Mser}) and (\ref{psiser}),
respectively, we shall rewrite the boundary conditions as 
\begin{equation}
\lim_{r\rightarrow \infty }M_{0}\left( r\right) =\mathcal{M,}  \label{M0}
\end{equation}
with vanishing  $M_{1}(\infty)$ and $\psi_{1}(\infty).$
%%%%%%%%%%%%%%%%%%%%%%%

Inserting Eq.~(\ref{sabg}) into (\ref{labg}) and makig use of Eq.~(\ref{postacF}) one has 
\begin{equation}
\mathcal{L}\left( F\right) =\frac{2Q^{2}}{r^{4}}\left( 1-\tanh ^{2}\frac{
Q^{2}}{2br}\right) .
\end{equation}
Subsequently expanding the right hand side of Eqs. (\ref{2nd_order}) with
respect to $\varepsilon $ and retaining the linear terms only yields 
\begin{equation}
8\pi T_{t}^{t}=8\pi T_{r}^{r}=-\frac{Q^{2}}{r^{4}}\left( 1-\tanh ^{2}\frac{
Q^{2}}{2b_{1}r}\right) -\varepsilon \frac{b_{2}Q^{4}}{b_{1}^{2}r^{5}}\left(
\cosh ^{-2}\frac{Q^{2}}{2b_{1}r}\,\tanh \frac{Q^{2}}{2b_{1}r}\right) .
\label{tep}
\end{equation}
The zeroth-order equation 
\begin{equation*}
M_{0}^{\prime }\left( r\right) =\frac{Q^{2}}{2r^{2}}\left( 1-\tanh ^{2}\frac{
Q^{2}}{2b_{1}r}\right)
\end{equation*}
can be easily integrated 
\begin{equation}
M_{0}\left( r\right) =C_{2}-b_{1}\tanh \frac{Q^{2}}{2b_{1}r},  \label{mm0}
\end{equation}
where $C_{2}$ is the integration constant. Making use of the condition (\ref
{M0}) gives $C_{2}=\mathcal{M}$. On the other hand, demanding of the
regularity of the line element as $r\rightarrow 0$ yields $b_{1}=\mathcal{M,}
$ and, consequently, $M_{0}\left( r\right) $ reads 
\begin{equation}
M_{0}\left( r\right) =\mathcal{M}\left( 1-\tanh \frac{Q^{2}}{2\mathcal{M}r}
\right) .
\label{eMzero}
\end{equation}
The zeroth-order solution has an interesting property, which, as we
shall see, is crucial in our subsequent analysis: $M_{0}(r)$ as well
as its derivatives vanish in the limit $r\to 0.$ Moreover, it should
be emphasized that although the integration constant and the free
parameter in the thus constructed solution are equal to the total mass
of the system as seen by a distant observer their status is different:
the interpretation of the former is an inevitable consequence of the
boundary conditions whereas the latter should be postulated.

For small values of $|Q|/\mathcal{M}$ as well as at great distances
the ABGB line element resembles that of Reissner-Nordstr\"om.
It can be easily seen by expanding the function $M_{0}(r)$
\begin{equation}
g_{tt}\,=-1+\frac{2\mathcal{M}}{r} - \frac{Q^2}{r^{2}}\,
+\,\frac{Q^{6}}{12\mathcal{M}^{2}r^{4}}\,+\,...
,
\end{equation} 
whereas for $r \to 0$ one has
\begin{equation}
f \sim 1 -\frac{4\mathcal{M}}{r}\exp\left(\frac{-Q^{2}}{\mathcal{M}r}\right).
\end{equation}
Noticeable differences appear for the configurations near the extremality
limit and in the internal region in the vicinity of the center. 
Indeed, for the ABGB solution $g_{tt}$ tends to $-1$ as $ r \to 0$ 
whereas the $(00)$ component of the metric tensor 
of the Reissner-Nordstr\"om solution diverges in that limit as $-r^{-2}.$   

From (\ref{pseq}) one concludes that the zeroth-order solution, 
$M_{0}\left(r\right) $, is sufficient to determine the function 
$\psi _{1}\left(r\right) .$ Indeed, substituting (\ref{eMzero}) 
into Eq. (\ref{pseq}) and making use of the condition (\ref{bound2}) 
one obtains 
\begin{eqnarray}
\psi _{1}\left( r\right) &=&\left[ \frac{\beta Q^{2}}{r^{4}}+\frac{\left(
2\alpha +\beta \right) Q^{6}}{2\mathcal{M}^{2} r^{6}}\tanh ^{2}\frac{Q^{2}}{2
\mathcal{M}r}-\frac{3\left( 2\alpha +\beta \right) Q^{4}}{\mathcal{M}r^{5}}
\tanh \frac{Q^{2}}{2\mathcal{M}r}\right] \cosh ^{-2}\frac{Q^{2}}{2\mathcal{M}
r}  \notag \\
&&-\frac{\left( 2\alpha +\beta \right) Q^{6}}{4\mathcal{M}^{2}r^{6}}\cosh
^{-4}\frac{Q^{2}}{2\mathcal{M}r}.
\label{psi_zeroo}
\end{eqnarray}
It could be easily demonstrated that that $\psi_{1}(r)$ vanishes at $r=0$
and inspection of Eq.~(\ref{pseq}) reveals similar behaviour of 
its derivatives.

The solution of the first-order equation 
\begin{equation}
\frac{2}{r^{2}}(M_{1}^{\prime }-S_{t}^{t})=\frac{b_{2}Q^{4}}{b_{1}^{2}r^{5}}
\left( \cosh ^{-2}\frac{Q^{2}}{2b_{1}r}\,\tanh \frac{Q^{2}}{2b_{1}r}\right) 
\label{f_o}
\end{equation}
is more complicated and when
combined with the appropriate boundary conditions it could be written as 
\begin{equation}
M_{1}\left( r\right) =\int_{\infty }^{r}S_{t}^{t}\left( r\right) dr+\frac{
b_{2}Q^{4}}{2\mathcal{M}^{2}}\int_{\infty }^{r}\frac{1}{r^{3}}\left( \cosh
^{-2}\frac{Q^{2}}{2\mathcal{M}r}\,\tanh \frac{Q^{2}}{2\mathcal{M}r}\right)
dr.  \label{eqM1}
\end{equation}
Let us consider the first integral in the right hand side of (\ref{eqM1})
first. The result can be expressed in terms of the hyperbolic functions and
polylogarithms $\mathrm{Li}_{i}\left( s\right) $. Indeed, utilizing formulas
collected in Appendix one can construct the solution which has the general
structure 
\begin{eqnarray}
M_{1}^{\left( 0\right) }\left( r\right) &=&-\int_{\infty
}^{r}S_{t}^{t}\left( r\right) dr=\frac{1}{\mathcal{M}}\sum_{i=0}^{1}
\sum_{j=0}^{2}\left( \alpha \widetilde{f}_{ij}^{\left( \alpha \right)
}+\beta \widetilde{f}_{ij}^{\left( \beta \right) }\right) \tanh ^{i}\dfrac{
Q^{2}}{2\mathcal{M}r}\sec \mathrm{h}^{2j}\dfrac{Q^{2}}{2\mathcal{M}r}  \notag
\\
&&+\frac{1}{\mathcal{M}}\sum_{i=1}^{6}\left( \alpha \widetilde{h}
_{i}^{\left( \alpha \right) }+\beta \widetilde{h}_{i}^{\left( \beta \right)
}\right) \mathrm{Li}_{i}\left( -\exp \left( -\frac{Q^{2}}{\mathcal{M}r}
\right) \right) -\mu .  \label{sol1}
\end{eqnarray}
Here $\mu $ is given by 
\begin{equation}
\mu =\frac{\pi ^{2}\mathcal{M}^{5}}{Q^{6}}\left[ \alpha \left( 8-\frac{31}{
315}\pi ^{4}\right) +\beta \left( \frac{8}{3}+\frac{7}{45}\pi ^{2}-\frac{31}{
630}\pi ^{4}\right) \right]\,=\,\frac{\pi^{2} \mathcal{M}^{5}}{Q^{6}}\sigma  \label{mu},
\end{equation}
and since the terms in the square brackets will appear frequently 
in our subsequent analyses we have singled them out and denoted by $\sigma .$  
 The functions $\tilde{f}$\ and $\tilde{h}$ are simple polynomials of 
$r^{-1};$ their actual form will not be displayed here as we shall readily 
rewrite them in a slightly modified form. It should be noted, however,
that a careful term-by-term analysis of Eq.~(\ref{sol1}) reveals 
that $M_{1}^{(0)}(r)$ approaches $-\mu$ as $r \to 0.$ Since for $b=b_{1}$
the functions $M_{1}(r)$ and $M_{1}^{(0)}$ coincide it is impossible to
obtain a regular solution at the center, and the remedy is to retain $b_{2}$
in the calculations.

Converting all hyperbolic functions appearing in the solution to the
exponents, introducing a new function $\xi $ defined as
\begin{equation}
\xi \left( r\right) =\exp \left( -\frac{Q^{2}}{\mathcal{M}r}\right)
\label{xi1}
\end{equation}
and finally making use of the elementary properties of the function $\rm{Li}_{0}(-\xi)$
(see Eq.~(\ref{AA}) of the Appendix)
%  \begin{equation}
%  \mathrm{Li}_{0}\left( -\xi \right)\,=\,-\frac{\xi}{1 + \xi},
%  \end{equation}
one has 
\begin{equation}
M_{1}^{\left( 0\right) }\left( r\right) =\sum_{i=0}^{6}\left( \alpha
f_{i}^{\left( \alpha \right) }+\beta f_{i}^{\left( \beta \right) }\right) 
\mathrm{Li}_{i}\left( -\xi \right) -\mu .  \label{tylda}
\end{equation}
The term $\mu $ has been singled out for convenience and $f_{i}^{\left(
\alpha \right) }$ and $f_{i}^{\left( \beta \right) }$ are given respectively by 
\begin{eqnarray}
&&f_{0}^{\left( \alpha \right) }=-\frac{48\mathcal{M}^{3}}{Q^{2}r^{2}}+\frac{
4\mathcal{M}Q^{2}\left( \xi -1\right) }{r^{4}\left( 1+\xi \right) ^{2}}-
\frac{16\mathcal{M}^{2}}{r^{3}\left( 1+\xi \right) }-\frac{4Q^{6}\left(
1-4\xi +\xi ^{2}\right) }{\mathcal{M}^{2}r^{5}\left( 1+\xi \right) ^{3}} 
\notag \\
&&-\frac{4Q^{4}\left[ \mathcal{M}\left( 1+86\xi -89\xi ^{2}\right)
+25r\left( \xi ^{2}-1\right) \right] }{5\mathcal{M}r^{5}\left( 1+\xi \right)
^{3}}+\frac{2Q^{6}\xi \left( 75-425\xi +125\xi ^{2}+\xi ^{3}\right) }{15
\mathcal{M}r^{6}\left( 1+\xi \right) ^{4}},
\end{eqnarray}
\begin{equation}
f_{1}^{\left( \alpha \right) }=\frac{4Q^{4}}{5r^{5}}-\frac{96\mathcal{M}^{4}
}{Q^{4}r},\qquad f_{2}^{\left( \alpha \right) }=\frac{4\mathcal{M}Q^{2}}{
r^{4}}-\frac{96\mathcal{M}^{5}}{Q^{6}},
\end{equation}
\begin{equation}
\frac{6r^{3}\mathcal{M}^{3}}{Q^{6}}f_{3}^{\left( \alpha \right) }=\frac{
2r^{2}\mathcal{M}^{2}}{Q^{4}}f_{4}^{\left( \alpha \right) }=\frac{r\mathcal{M
}}{Q^{2}}f_{5}^{\left( \alpha \right) }=f_{6}^{\left( \alpha \right) }=\frac{
96\mathcal{M}^{5}}{Q^{6}}
\end{equation}
and
\begin{eqnarray}
f_{0}^{\left( \beta \right) } &=&-\frac{16\mathcal{M}^{3}}{Q^{2}r^{2}}-\frac{
16\mathcal{M}^{2}}{3r^{3}\left( 1+\xi \right) }-\frac{2Q^{2}\left[ 12r\left(
1+\xi \right) +\mathcal{M}\left( 3-36\xi +\xi ^{2}\right) \right] }{
3r^{4}\left( 1+\xi \right) ^{2}}  \notag \\
&&-\frac{2Q^{6}\left( 1-4\xi +\xi ^{2}\right) }{\mathcal{M}^{2}r^{5}\left(
1+\xi \right) ^{3}}+\frac{Q^{6}\xi \left( 75-425\xi +125\xi ^{2}+\xi
^{3}\right) }{15\mathcal{M}r^{6}\left( 1+\xi \right) ^{4}}  \notag \\
&&-\frac{2Q^{4}\left[ \mathcal{M}\left( 1+96\xi -109\xi ^{2}\right)
+30r\left( \xi ^{2}-1\right) \right] }{5\mathcal{M}r^{5}\left( 1+\xi \right)
^{3}},
\end{eqnarray}
\begin{equation}
f_{1}^{\left( \beta \right) }=\frac{2Q^{4}}{5r^{5}}-\frac{8\mathcal{M}^{2}}{
3r^{3}}-\frac{32\mathcal{M}^{4}}{Q^{4}r},\qquad f_{2}^{\left( \beta \right)
}=-\frac{32\mathcal{M}^{5}}{Q^{6}}+\frac{2\mathcal{M}Q^{2}}{r^{4}}-\frac{8
\mathcal{M}^{3}}{Q^{2}r^{2}},
\end{equation}
\begin{equation}
f_{3}^{\left( \beta \right) }=\frac{8\mathcal{M}^{2}}{r^{3}}-\frac{16
\mathcal{M}^{4}}{Q^{4}r},\qquad f_{4}^{\left( \beta \right) }=-\frac{16
\mathcal{M}^{5}}{Q^{6}}+\frac{24\mathcal{M}^{3}}{Q^{2}r^{2}},
\end{equation}
\begin{equation}
\frac{r\mathcal{M}}{Q^{2}}f_{5}^{\left( \beta \right) }=f_{6}^{\left( \beta
\right) }=\frac{48\mathcal{M}^{5}}{Q^{6}}.
\end{equation}
The second quadrature in the right hand side of Eq. (\ref{eqM1}) is
elementary and gives 
\begin{equation}
M_{1}^{\left( 1\right) }=\frac{b_{2}Q^{4}}{2\mathcal{M}^{2}}\int_{\infty
}^{r}\frac{1}{r^{3}}\left( \cosh ^{-2}\frac{Q^{2}}{2\mathcal{M}r}\,\tanh 
\frac{Q^{2}}{2\mathcal{M}r}\right) dr=\frac{b_{2}Q^{2}}{2\mathcal{M}r}\cosh
^{-2}\frac{Q^{2}}{2\mathcal{M}r}-b_{2}\tanh \frac{Q^{2}}{2\mathcal{M}r}
\label{M_1_1}
\end{equation}
or equivalently 
\begin{equation}
M_{1}^{(1)}\,=\,b_{2}\left[\frac{1-\xi}{\xi} 
- \frac{2Q^{2}}{\mathcal{M}r (1+\xi)} \right] {\rm Li}_{0}(-\xi) .
\end{equation}
Now, in order to obtain a regular solution at the center one has to adjust
suitably the parameter $b_{2}.$ To accomplish this let us observe that 
$M_{1}^{\left(1\right) }\left( 0\right) =-b_{2}.$ 
Further, taking the results of the discussion
presented below Eq.~(\ref{mu}) into account one finds 
\begin{equation}
b_{2}=-\mathcal{\mu} .
\end{equation}
%%%%%%%%%%%%%%%%%%%%%%%%%
%%%%%  Eq. (\ref{sol1})  
%%%%%%%%%%%%%%%%%%%%%%%%%
It could be easily seen that with such a choice of $b_{2}$ the function 
$M_{1}\left(r\right) $ vanishes at the center and as $r\rightarrow \infty $
as required. 

It should be noted that the representation of the thus obtained line element
is by no means unique. For example one can express the result in terms
of $\rm{Li}_{n}(-1/\xi)$ rather than  $\rm{Li}_{n}(- \xi)$ employing
identity~\cite{Levin1}
\begin{equation}
{\rm Li}_{n}(-\xi) + (-1)^{n} {\rm Li}_{n}(-1/ \xi)\,=\,-\frac{1}{n!} \ln^{n} \xi +
2 \sum_{r=1}^{[n/2]} \frac{\ln^{n-2r}\xi}{(n-2r)!}\, {\rm Li}_{2 r}(-1) ,
\end{equation}
where $[n/2]$ is the greatest integer contained in $n/2$ and the constants 
${\rm Li}_{2 r}(-1)$ are related to the Bernoulli numbers, $B_{2 r},$  
according to the formula
\begin{equation}
{\rm Li}_{2 r}(-1)\,=\frac{2^{2r-1}}{(2r)!} B_{2 r} \pi^{2 r}.
\end{equation}
Regardless of the chosen representation, after imposing boundary
conditions, both solutions are, of course, equivalent. 

As the ABGB geometry reduces to the Schwarzschild solution in the limit $Q=0,$
it is the charge, no matter how small, that secures regularity. The vanishing 
of the charge leads therefore to dramatic changes in the geometry of the black
hole interior.
On the other hand, the terms calculated to ${\mathcal O}(Q^{2})$  for the ABGB spacetime 
coincide with the Reissner-Nordstr\"om solution.
Let us consider a series expansion of the function $M_{1}^{\left( 1\right)
}\left( r\right) $ for small $q =|Q|/\mathcal{M}.$ After some algebra one has
\begin{eqnarray}
M_{1}&=&\left(\frac{2}{x^{3}} - \frac{3}{x^{4}} \right)\frac{\beta}{\mathcal{M}}q^{2} + 
\frac{6\beta}{5\mathcal{M}}q^{4} - \left[\left( \frac{3}{x^{5}} 
-\frac{11}{2 x^{6}} \right)\alpha 
+ \left(\frac{9}{4 x^{5}} - \frac{4}{x^{6}} \right)\beta\right] 
\frac{q^{6}}{\mathcal{M}}\nonumber \\
&-&\left(\frac{5}{2}\alpha + \frac{13}{7}\beta \right)\frac{q^{8}}{\mathcal{M}x^{7}}\,+\,...
-b_{2} \left(\frac{q^{6}}{12 x^{3}} - \frac{q^{10}}{60 x^{5}} 
+ \frac{17q^{14}}{6720 x^{7}} \right) + ...,
             \label{exp_M1}
\end{eqnarray}
where $x=r/\mathcal{M}.$
It could be easily checked that for the regular line element 
the leading term of the expansion is proportional to $b_{2}Q^{6}.$ 
Since the coefficient $b_{2}$ is proportional to $Q^{-6}$ 
the constructed line element does not approach the Schwarzschild
solution in the limit $Q \to 0$. Indeed, simple calculations yield
\begin{equation}
ds^{2}=-\left( 1-\frac{2\mathcal{M}}{r}-\frac{2\mathcal{M}^{2}}{r^{4}}
k\right) dt^{2}
+\left[\left( 1-\frac{2\mathcal{M}}{r}\right)^{-1}\,+\,
\left( 1-\frac{2\mathcal{M}}{r}\right)^{-2}
\frac{2\mathcal{M}^{2}}{r^{4}
}k \right]dr^{2}
+r^{2}d\Omega ^{2},  \label{limitt}
\end{equation}
where $k$ \ is given by 
\begin{equation}
k=\frac{\pi ^{2}}{12}\sigma \approx 5.781\alpha -0.486\beta  .
\end{equation}
Consequently, one has either Schwarzschild asymptotic of a singular 
line element with $b_{2} =0$ ($k =0$) or a regular line element with non-Schwarzschild 
$Q = 0$ limit for $b_{2} = - \mu.$ 
It is possible, of course, to accept other values of the parameter $b_{2}$
but it seems that they are of lesser importance.

The approximate location of the event horizon of the line element (\ref
{limitt}) lies near its Schwarzschild value and is approximately given by
\begin{equation}
r_{+}\approx 2\mathcal{M}\left( 1+\frac{k}{8\mathcal{M}^{2}}\right) .
\end{equation}
As is well-known, the Hawking temperature, $T_{H},$ can be easily 
calculated from the metric tensor without referring to the field equations. 
The standard by now method of obtaining $T_{H}$ relies on the Wick rotation. 
The Euclidean line element has no conical singularity provided 
the time coordinate is periodic with a period $P$ given by 
\begin{equation}
P=4\pi \lim_{r\rightarrow r_{+}}\left( g_{tt}g_{rr}\right) ^{1/2}\left( 
\frac{d}{dr}g_{tt}\right) ^{-1} .
                                    \label{period}
\end{equation}
Its reciprocal is identified with the black hole temperature, 
which, in the case in hand, reads 
\begin{equation}
T_{H}\,=\,\frac{1}{8\pi\mathcal{M}}\left(1 + \frac{k}{4\mathcal{M}^{2}} \right) .
                                    \label{th}
\end{equation}
Note that for $\alpha$ and $\beta$ satisfying 
\begin{equation}
\alpha = \frac{1680 + 98\pi^{2} - 31\pi^{4}}{2 (31 \pi^{4} - 2520)}\beta,
\end{equation}
both (\ref{limitt}) and (\ref{th}) reduce to their Schwarzschild counterparts.
 
Finally, we shall investigate the important issue of regularity. 
However, before we proceed further a few words of comments are in
order. It should be emphasized that we have constructed a linearized
solution to the coupled system of equations of quadratic gravity and 
nonlinear electrodynamics only. Consequently, when calculating 
the curvature invariants we are to restrict ourselves to $O(\varepsilon)$ 
terms. 
As it has been stressed earlier, the regularity of the line 
element does not necessarily entail the regularity of the underlying geometry: 
the latter requires various curvature invariants to be regular in 
the interesting region. First, let us concentrate on the Kretschmann scalar, $K.$
It could be easily shown that for the constructed line element
it consists of terms involving products of 
the functions $M_{0}(r),$ $M_{1}(r),$  $\psi_{1}(r)$ and their derivatives. Indeed,
after some algebra one obtains $K = K_{0} + \varepsilon \Delta K,$
where $K_{0}$ is the Kretschmann scalar of the ABGB spacetime
\begin{equation}
K_{0} = \frac{48 M_{0}^{2}}{r^{6}} +\frac{16 M_{0}^{\prime\prime}M_{0}}{r^{4}} 
+\frac{32 {M_{0}^{\prime}}^{2}}{r^{4}} - \frac{64 M_{0}^{\prime} M_{0}}{r^{5}}
+\frac{4 {M_{0}^{\prime\prime}}^{2}}{r^{2}} - 
\frac{16 M_{0}^{\prime\prime}M_{0}^{\prime}}{r^{3}}
\end{equation}
and $\Delta K$ is given by
\begin{eqnarray}
\Delta K &=&\left( \frac{128 M_{0} M_{0}^{\prime}}{r^{4}}+\frac{16 M_{0}}{r^{4}}
-\frac{80 M_{0}^{2}}{r^{5}} 
-\frac{16 M_{0}^{\prime}}{r^{3}} - \frac{24 M_{0} M_{0}^{\prime\prime}}{r^{3}} + 
\frac{24 M_{0}^{\prime} M_{0}^{\prime\prime}}{r^{2}} -\frac{48 {M_{0}^{\prime}}^{2}}{r^{3}}
\right)\psi_{1}^{\prime}\nonumber \\
&+&
\left(-\frac{8 M_{0}^{\prime\prime}}{r} + 
\frac{16 M_{0} M_{0}^{\prime\prime}}{r^{2}}+\frac{32 M_{0}^{2}}{r^{4}} 
+ \frac{16 M_{0}^{\prime}}{r^{2}} - \frac{32 M_{0}M_{0}^{\prime}}{r^{3}} 
- \frac{16 M_{0}}{r^{3}}\right)
\psi_{1}^{\prime\prime}\nonumber\\
&+&\left(-\frac{16 M_{0}^{\prime\prime}}{r^{4}} 
-
\frac{96 M_{0}}{r^{6}} + \frac{64 M_{0}^{\prime}}{r^{5}} \right)M_{1}
+
\left(\frac{16 M_{0}^{\prime}}{r^{3}} -\frac{16 M_{0}}{r^{4}} 
-\frac{8 M_{0}^{\prime\prime}}{r^{2}} \right)M_{1}^{\prime\prime}\nonumber \\
&+&
\left(\frac{16 M_{0}^{\prime\prime}}{r^{3}} - 
\frac{64 M_{0}^{\prime}}{r^{4}} +\frac{64 M_{0}}{r^{5}} \right) M_{1}^{\prime} .
\end{eqnarray}
The regularity of $K_{0}$ has been demonstrated in Ref~\cite{ABG};
in order to establish regularity of $\Delta K$
one has to establish regularity of its building blocks.
To analyse behaviour of the $k-$th derivative of $M_{1}$
we shall employ the first order equation. From Eq.~(\ref{1sta}) one clearly sees
that all derivatives of $S_{t}^{t}$ vanish as $r\to 0.$ 
It is simply because $k-$th derivative of $M_{0}$  has the asymptotic
form 
\begin{equation}
\frac{d^{k}M_{0}}{dr^{k}} \sim 
\sum_{i=k+1}^{2k}\frac{c_{i}}{r^{i}} \exp\left(-\frac{Q^{2}}{\mathcal{M}r}\right),
\end{equation}
as $r \to 0,$
where the coefficients $c_{i}$ depend on $\mathcal{M}$ and $Q.$ 
It follows then that $k$-th derivative behaves as
$r^{-2k} \exp(-Q^{2}/\mathcal{M}r)$  near the center.
One encounters a similar behaviour in the second 
term in the right hand side of Eq.~(\ref{eqM1}).
Pulling this all together one concludes that the Kretschmann scalar is indeed regular
at the center. Moreover, one can easily establish the regularity 
of $R^{2}$ and $R_{ab} R^{ab}.$ Similar arguments
can be used in demonstration that higher invariants constructed from the curvature
are also regular.

\section{Geometry}
 \label{sec4a}
Generally speaking a black hole belongs to one of the two distinct
classes: it may be either extremal (when horizon has at least twofold
degeneracy) or nonextremal. In the former case the family of
ultraextremal black holes, i. e. configurations with triple (or even
higher) degeneracy can be singled out~\cite{dirty_bh,semi}.
Since we have restricted ourselves to the case of vanishing cosmological
constant, we expect that the ABGB black hole possesses at most two horizons. 
(Generalization of the ABGB solution to $\Lambda \neq 0$ case is straightforward.)
Indeed, depending on $q=|Q|/\mathcal{M}$ the ABGB spacetime has two, one, or has no
horizons at all. Simple calculations indicate that for $q<q_{c},$ where
$q_{c}$ is a critical value of $q,$
it has both the inner and outer horizon, whereas for $q>q_{c\text{ }
}$ the metric potential $g_{tt}(r)$ has no real roots for $r \geq 0.$ For 
$q=q_{c\text{ }}$ the event and inner horizons
merge and the configuration becomes extremal. 
It should be noted that the extremal ABGB geometry
is distinct from, say, the geometry of the extremal 
Reissner-Nordstr\"om solution.
Indeed, because of the regularity of the former, the configurations with 
$q>q_{c}$ are not forbidden by a cosmic censor.

\subsection{Nonextreme black holes}
 \label{sec4}
In order to determine the location of the inner and outer horizon
we shall analyse equation $
g_{tt}\left( r\right) =0$ and construct the iterative solution restricting
ourselves
to the terms linear in $\varepsilon .$ As the\ solution can be written as 
\begin{equation}
r_{\pm }=r_{0}^{\left( \pm \right) }+\varepsilon r_{1}^{\left( \pm \right) },
\end{equation}
one has to solve the simple system of two equations. First of them 
\begin{equation}
1-\frac{2M_{0}\left( r_{0}^{\left( \pm \right) }\right) }{r_{0}^{\left( \pm
\right) }}=0  \label{hor+1}
\end{equation}
admits solutions expressible in terms of the Lambert special functions.
Indeed, it has been demonstrated in Ref. \cite{Kocio1} that 
$x_{0}^{\left( \pm\right) }=r_{0}^{\left( \pm \right) }\slash\mathcal{M}$ 
are given by 
\begin{equation}
x_{0}^{\left( \pm \right) }=-\frac{4q^{2}}{4W_{\pm }\left( -\frac{q^{2}}{4}%
\exp \left( \frac{q^{2}}{4}\right) \right) -q^{2}},  \label{x1}
\end{equation}
where $W_{+}\left( s\right) \equiv W(0,s)$ and $W_{-}\left( s\right) \equiv
W(-1,s)$ are the only real branches of the Lambert functions~\cite{Knuth}. 
On the other hand, $r_{1}$ satisfies algebraic equation
which can be easily solved. After some manipulations one has 
\begin{equation}
r_{1}^{\left( \pm \right) }=\frac{2M_{1}\left( r_{0}^{\left( \pm \right)
}\right) }{1-2M_{0}^{\prime }\left( r_{0}^{\left( \pm \right) }\right) }.
\label{hor+2}
\end{equation}
Now, inserting the zeroth-order solutions (\ref{x1}) 
into Eq.~(\ref{hor+2}) one can easily determine $r_{1}^{\left( \pm \right) },$ 
and the result can be further simplified with the substitution 
\begin{equation}
x_{0}=\frac{4\xi }{\xi +1},
\end{equation}
where $\xi $ (defined as in Eq. (\ref{xi1})) \ is calculated at $x=x_{0}.$
Specifically, denominator in the right hand side of Eq. (\ref{hor+2}) 
reduces to the simple expression:
\begin{equation}
1-2M_{0}^{\prime }\left( r_{0}^{\left( \pm \right) }\right) =\frac{4\xi
-q^{2}}{4\xi }.
\end{equation}
To avoid unnecessary proliferation of long formulas we shall not present
the results for $r_{1}^{\left( \pm \right) }$ here as they could be readily
obtained by a direct substitution of the equation (\ref{x1}) and 
the expression describing 
$M_{1}$ at $ r^{(\pm)}_{0} $ into Eq.~(\ref{hor+2}).

Although the complexity of the expression describing $r_{+}$
makes its direct analytical application rather intricate,
one can obtain interesting and important information analyzing 
limiting cases. Below we derive expansion valid for 
small $q.$ The extremality limit will be analyzed in the next subsection.
Expanding $r_{+}$ and collecting the terms with like powers of $q,$
after massive simplifications, one has
\begin{eqnarray}
r_{+} &=& 2\mathcal{M} - \frac{q^{2}}{2}\mathcal{M} -
\frac{q^{4}}{8}\mathcal{M} - \frac{5q^{6}}{96}\mathcal{M} +\, ...\,
+ \frac{1}{4\mathcal{M}}k +\left(\frac{1}{4\mathcal{M}}k 
+\frac{\beta}{8\mathcal{M}} \right)q^{2}
\nonumber \\
&+& \left(\frac{71}{320 \mathcal{M}}k+\frac{17}{160\mathcal{M}}\beta \right)q^{4}
+ \left(\frac{123}{640\mathcal{M}}k + 
\frac{47}{640\mathcal{M}}\beta -\frac{1}{64\mathcal{M}}\alpha \right)q^{6} + ...
                                           \label{rozw1}
\end{eqnarray}
Analogous expression for the event horizon of the Reissner-Nordstr\"om 
solution in quadratic gravity reads
\begin{eqnarray}
r_{+} &=& 2\mathcal{M} - \frac{1}{2}\mathcal{M}q^{2} - \frac{1}{8}\mathcal{M}q^{4}
- \frac{1}{16}\mathcal{M}q^{6} + ... \nonumber \\
&+& \left(\frac{1}{8\mathcal{M}}q^{2} + \frac{17}{160\mathcal{M}}q^{4}
+ \frac{57}{640\mathcal{M}}q^{6} + ...  \right)\beta .
                       \label{rozw2}
\end{eqnarray}
Inspection of (\ref{rozw1}) and (\ref{rozw2}) shows that for $b_{2} = 0$
(or, equivalently, $k =0$)  both expansion coincide up to $q^{4}$ terms.

One of the most important characteristics of the black hole is
its temperature. To investigate how it is modified by the quadratic terms,
we employ a general expression~(\ref{period}), 
which, in the present context, can be rewritten in the form
\begin{equation}
T_{H} = \frac{1}{4\pi} \lim_{r \to \rp} e^{-\varepsilon \psi_{1}} 
\left( \frac{1}{r}- \frac{2\varepsilon}{r}S^{t}_{t} + 8\pi r T^{t}_{t} \right).
\end{equation} 
Further, expanding $T_{H},$ in the auxiliary parameter, 
collecting the terms with like powers of 
$\varepsilon$ and linearizing the thus obtained result one has
\begin{equation}
T_{H} = T_{0} + \varepsilon \Delta T ,
\end{equation} 
where $T_{0}$ coincides with the temperature of the nonextremal
ABGB black hole
\begin{equation}
T_{0} = \frac{1}{4\pi}\left[ \frac{1}{r_{0}} - 
\frac{Q^{2}}{\mathcal{M} r_{0}^{2}}\left( 1 - \frac{r_{0}}{4 \mathcal{M}} \right)\right]
\end{equation}
and $\Delta T$ is given by
\begin{equation}
\Delta T = r_{1}\left(2 r_{0}
\frac{d T^{t}_{t}}{dr}_{| r_{0}}  -\frac{1}{2 \pi  r_{0}^{2}} +
\frac{T_{0}}{r_{0}} \right) - \frac{1}{2 \pi r_{0}} S_{t}^{t} + 
T_{0} \psi_{1} (r_{0}) +2 r_{0} T^{t}_{t}.
\end{equation}
Here $r_{0} =  r_{0}^{(+)},$ whereas $T^{t}_{(0)t}$ and $T_{(1)t}^{t} $
are given by the first and the second term in the right hand side of 
Eq.~(\ref{tep}), respectively.

Of course, the general expression for temperature 
is also too complicated to be displayed here.
We shall, therefore, calculate its limiting behaviour for small $q$
precisely in the same manner as it has been done 
with the location of the event horizon $r_{+}.$
As the temperature of the extremal configuration is zero,
this condition can be used in establishing relation between $Q$
and $\mathcal{M}.$ This will be done in the next subsection.
The expansion of the temperature of the nonextremal black hole reads
\begin{eqnarray}
T_{H} &=& \frac{1}{8\pi \mathcal{M}} - \frac{q^{4}}{128\pi\mathcal{M}} - 
\frac{5 q^{6}}{768\pi \mathcal{M}} + ...\nonumber\\
&+&\frac{k}{32\pi \mathcal{M}^{3}} + \frac{q^{2}}{64\pi \mathcal{M}^{3}}(3 k -\beta) 
+\frac{q^{4}}{80\pi\mathcal{M}^{3}}\left(\frac{33}{8}k - \beta \right) + ...
\end{eqnarray}
whereas for the Reissne-Norstr\"om geometry in quadratic gravity one has
\begin{equation}
T_{H} = \frac{1}{8\pi\mathcal{M}} - \frac{q^{4}}{128\pi\mathcal{M}} -\frac{q^{6}}{128\pi\mathcal{M}}
+\, ... \,+\left( \frac{q^{4}}{320 \mathcal{M}}  + \frac{3 q^{6}}{512 \mathcal{M}}\,+ ...\right)\beta .
\end{equation}

%%%%%%%%%%

\subsection{Extreme black holes}
 \label{sec5}

In this section we shall investigate the important issue of extremal black
holes. When $r_{+}$ and $r_{-}$ of the classical ABGB black hole merge into
the one degenerate horizon one has the extremal configuration characterized
by vanishing surface gravity (Hawking temperature). The geometry of the vicinity of the
degenerate horizon belongs to the 
${\rm AdS}_{2}\times {\rm S}^{2}$ class with different modules
of curvatures of subspaces
\cite{Kocio3}. One of the peculiarities of the
general extremal solution is infinite proper distance between two points,
one of which resides on the event horizon.
%%%%% !!!!!!!! %%%%%%
The degeneracy means that in the vicinity 
of the horizon the leading behaviour of the metric potential 
$g_{tt}$ is of the form
$g_{tt}\sim \left( r-r_{+}\right) ^{2},$ or equivalently 
\begin{equation}
g_{tt}\left( r_{+}\right) =g_{tt}^{\prime }\left( r_{+}\right) =0.
\label{eqqss}
\end{equation}
Combining these equations one obtains simple algebraic equation
that could be easily solved
\begin{equation}
x_{+} = \frac{4 q_{c}^{2}}{4-q_{c}^{2}}.
\end{equation} 
It can be demonstrated that the parameters 
$q$ and $x_{x}$ 
of the classical extremal ABGB
black hole can be expressed in terms of the principal 
branch of the Lambert function
evaluated at $e^{-1}.$
Employing definition of $W_{+}$ one has
\begin{equation}
q_{c}=2\sqrt{W_{+}\left( e^{-1}\right) }\approx 1.0554, 
\end{equation}
whereas the location of the degenerate horizon is given by
\begin{equation}
x_{c}=\frac{4W_{+}\left( e^{-1}\right) }{1+W_{+}\left( e^{-1}\right) }\approx 0.8712 .
\end{equation}
Here and below $x$ (with or without sub or superscripts) denotes appropriate
radial coordinate divided by $\mathcal{M}.$

Now, we shall examine modifications caused by quadratic gravity. 
Lengthy but routine calculations show that the extremal canfiguration
is possible for
\begin{equation}
q^{2}=q_{c}^{2} + \delta ,
\end{equation}
where the correction $\delta$ is given  by
\begin{equation}
\delta =\frac{1}{\mathcal{M}^{2}}\sum_{i=0}^{6}\left( \alpha h_{i}^{\left(
\alpha \right) }+\beta h_{i}^{\left( \beta \right) }\right) \mathrm{Li}%
_{i}\left( w\right) \,+\,\frac{2(3w-1)}{\mathcal{M}}b_{2}
\label{delta}
\end{equation}
with
\begin{equation}
h_{0}^{\left( \alpha \right) }=-\frac{(1+w)^{3}}{240 w^{3}}
\left( 516-69w+11w^{2}-5w^{3}-w^{4}\right)
- \frac{(1+w)^{2}\pi^{2}}{10080w^{4}}\left(31\pi^{2}-2520 \right) ,
\end{equation}
\begin{equation}
h_{1}^{\left( \alpha \right) }=-\frac{\left( 1+w\right) ^{2}}{40w^{3}}\left(
119-4 w-6 w^{2}-4 w^{3}-w^{4}\right) ,
\end{equation}
\begin{equation}
h_{2}^{\left( \alpha \right) }=-\frac{1}{8w^{3}}\left(1+w \right)\left(
23-4 w-6 w^{2}-4 w^{3}-w^{4}\right) ,
\end{equation}
\begin{equation}
\frac{2}{\left( 1+w\right) ^{3}}h_{3}^{\left( \alpha \right) }=\frac{2}{%
\left( 1+w\right) ^{2}}h_{4}^{\left( \alpha \right) }=\frac{1}{\left(
1+w\right) }h_{5}^{\left( \alpha \right) }=h_{6}^{\left( \alpha \right) }=%
\frac{3\left( 1+w\right) }{w^{3}},
\end{equation}
%%%%%%%%%%%%%%%%%%%%%%%%%%%%%%%%%%%%%%%%5
\begin{equation}
h_{0}^{\left( \beta \right) }=-\frac{(1+w)^{3}}{480w^{3}}
\left(356+21w+21w^{2}-5w^{3}-w^{4} \right)
+ \frac{(1+w)^{2}\pi^{2}}{20160w^{4}}\left(1680 + 98\pi^{2} - 31 \pi^{4} \right) ,
\end{equation}
\begin{equation}
h_{1}^{\left( \beta \right) }=-\frac{\left( 1+w\right) ^{2}}{240w^{3}}\left(
257+28w+2w^{2}-12w^{3}-3w^{4}\right) ,
\end{equation}
\begin{equation}
h_{2}^{\left( \beta \right) }=-\frac{1}{16w^{3}}\left(1+w\right)\left(
19+4w+2w^{2}-4w^{3}-w^{4}\right) ,
\end{equation}
\begin{equation}
h_{3}^{\left( \beta \right) }=-\frac{1}{4w^{3}}\left( 1+w \right)^{2}\left(
1-2w-w^{2}\right) ,\qquad h_{4}^{\left( \beta \right) }=\frac{1}{
4w^{3}}\left(1+w \right)\left( 1+6w+3w^{2}\right) ,
\end{equation}
\begin{equation}
\frac{1}{1+w}h_{5}^{\left( \beta \right) }=h_{6}^{\left( \beta \right) }=
\frac{3\left( 1+w\right) }{2w^{3}},
\end{equation}
and 
\begin{equation}
w=W_{+}\left( e^{-1}\right) .
\end{equation}
%%%%%%%%%%%%%%%%%%%%%%%%%%%%%%%%%%%%%%%%%%%%%%%%%%
The term containing $b_{2} (=- \mu),$ i. e., the last term
in right hand side of Eq.~(\ref{delta}) has been singled out for convenience.
For the extreme black hole $\mu$ reads
\begin{equation}
\mathcal{\mu }=\frac{\pi ^{2}}{64\mathcal{M}w^{3}}\sigma,
\end{equation}
where $\sigma$ is defined through the relation (\ref{mu}).
The location of the degenerate horizon is given by
\begin{equation}
x=x_{c}+ \Delta, 
\end{equation}
where $ \Delta$ can be compactly expressed in the form
%%%%%%%%%%%%%%%%%%%%% DELTA %%%%%%%%%%%%%%%%%%%%%%
\begin{equation}
\Delta =\frac{\delta w}{\left( 1+w\right) ^{2}}+
\frac{2\alpha+\beta}{16w\mathcal{M}^{2}}(w+3)(w-1)
-\frac{4w(w-1)}{\mathcal{M}(1+w)^{2}}b_{2},
\end{equation}
with $\delta$ given by Eq.~(\ref{delta}).
As both $\delta$ and $\Delta$  depend on the particular 
value of the Lambert function at $e^{-1}$ 
one can easily calculate their numerical values.
Indeed, simple calculations yield
\begin{equation}
\delta = -\frac{0.81451}{\mathcal{M}^{2}}\beta-\frac{3.67845}{\mathcal{M}^{2}}\alpha
\end{equation}
and
\begin{equation}
\Delta = 
\frac{1.40646}{\mathcal{M}^{2}}\beta + \frac{3.88206}{\mathcal{M}^{2}}\alpha ,
\end{equation}
for $b_{2} = - \mu,$ whereas with $b_{2} =0$ one has
\begin{equation}
\delta = -\frac{0.57553}{\mathcal{M}^{2}}\beta-\frac{0.05121}{\mathcal{M}^{2}}\alpha
\end{equation}
and
\begin{equation}
\Delta = 
-\frac{0.43288}{\mathcal{M}^{2}}\beta - \frac{1.05314}{\mathcal{M}^{2}}\alpha .
\end{equation}
Note that for particular choices of the coupling parameters it is possible
to reduce either $x_{+}$ or $q_{c}$ to its general relativistic
values.
Obtained corrections can be contrasted with the analogous results 
calculated for the extreme Reissner-Nordstr\"om black hole in quadratic gravity:
\begin{equation}
q_{c}^{2} = 1 + \frac{2}{5\mathcal{M}^{2}}\beta
\end{equation}
and 
\begin{equation}
x_{+} = 1 + \frac{1}{5\mathcal{M}^{2}}\beta.
\end{equation}

\section{Conclusion and summary}
\label{sec6}

In this paper we have constructed perturbative solutions describing
spherically symmetric and static black holes to the coupled equations
of fourth-order gravity with the source term given by the stress-
energy tensor of the nonlinear electrodynamics. The Lagrangian of the
nonlinear field is a natural generalization of the Bronnikov's
Lagrangian. Because of technical complexities we have restricted
ourselves to the first-order corrections, i. e. the terms proportional
to $\varepsilon.$ The obtained line element is parametrized by two
integration constants and free parameters. Integration constants are
related to the (magnetic) charge and a total mass of the system as
seen by a distant observer whereas the free parameters are adjusted to
make the resulting solution regular at the center. The regularity
heavily relies on the form of the zeroth-order solution which
coincides with the solution of the ABGB-type and the special
properties of the polylogarithms ${\rm Li}_{i}.$ The metric potentials
thus computed enabled construction of the basic characteristics of the
geometry and its asymptotics: location of the inner and outer horizon,
Hawking temperature and the relation between $Q$ and $\mathcal{M}$ for
the extremal configurations.

If one is interested in the solution possessing Schwarzschild
asymptotics as $Q\to 0$ or intend to study the external region 
without any relations to the issue of regularity  it suffices to put
$b_{2} =0$ throughout the paper. We intentionally presented basic
formulas in a form allowing for a different choices of the parameter
$b_{2}.$ It should be stressed that the only choice leading to
regularity of the solution at the center is $b_{2}\,=\,-\mu.$

The calculations and results presented in this paper suggest some 
generalizations and obvious extensions. First, one may attempt 
to go beyond first order calculations in order to establish full 
regularity of the solution. Of course it would be interesting and desirable, 
with all conceptual limitations of the method, to demonstrate it in a
nonperturbative way. However, the technical complexities may be a real
obstacle in this regard. Further, it is possible to analyse behaviour
of the test quantum fields in the ABGB background with the special
emphasis put on the back reaction on the metric. Indeed, since the
general expression describing the stress-energy tensor of quantized
massive scalar, spinor and vector fields in the large mass limit is
known~\cite{Matyjasek:1999an,Kocio1} it is possible to investigate the
resultant quantum-corrected geometry even in the vicinity of the
center of the black hole. 
Recently such a programme has been carried out in the case 
of the quantum-corrected Reissner-Nordstr\"om
black holes  (see Refs.~\cite{jirinek01b,jirinek03b} and
the references cited therein).
Finally, it should be noted that another
important choice of the boundary conditions
\begin{eqnarray}
M_{i}(\rp)\,=\,\begin{cases}
           \frac{\rp}{2} & \text{if $i=0$},\\
            0             & \text{if $i\geq 1$}
               \end{cases}
\end{eqnarray}
which is, in turn, related to the horizon defined mass of the black
hole is not considered in this paper. Here we remark only that
parametrizing solution by the exact location of the event horizon and
the charge is quite natural and certainly deserves further study. 
A lesson that follows from investigations of the corrected Reissner-Nordstr\"om
solution is that some relations, as for example these
for the extremal configurations, are simpler and physically more transparent.
We intend to return to these problems elsewhere.

\appendix* 
\section{}

In this appendix we sketch the method of evaluating the
indefinite integrals appearing in the solution of the equation (\ref{f_o}). 
After substitution of auxiliary variable $u=x^{-1}$ we have to find an
expression for the integral of the form: 
\begin{equation}
\int u^{p} (\tanh u)^{s} du,
\end{equation}
with natural $p, s$. We show that when $s>1$ the above integral can be
reduced to the integral with $s=1$ and then we will express the latter by a
sum containing polylogarithm functions. At the beginning we use the formulas
(1.4.22.3) and (1.4.22.4) from the very extensive tables of integrals~\cite
{Prud}:

\begin{equation}
\int u^{p} (\tanh u)^{2n} du = \frac{1}{p+1}u^{p+1}
+\sum_{k=0}^{n-1}(-1)^{n+k}\binom{n}{k}\int\frac{u^{p}}{(\cosh u)^{2n-2k}}
du,
\end{equation}
\begin{eqnarray}
\int u^{p} (\tanh u)^{2n+1} du &=& \int u^{p} \tanh u\,du 
+\sum_{k=0}^{n-1}
\frac{(-1)^{n+k}}{2n-2k}\binom{n}{k} \left[-\frac{u^{p}}{(\cosh u)^{2n-2k}}\right.
\nonumber \\
&&\left. +\, p\int\frac{u^{p-1}}{(\cosh u)^{2n-2k}} du\right].
\end{eqnarray}
Evaluation of integrals containing even powers $2n-2k>2$ of sech$\,$
function can be further reduced to the case with the square of ${\rm sech} $ in the
integrand by repeated use of the formula (1.4.24.1) of~\cite{Prud}: 
\begin{eqnarray}
\int \frac{u^{p}}{(\cosh u)^{q}} du &= & \frac{pu^{p-1}}{(q-1)(q-2)(\cosh
u)^{q-2}} + \frac{u^{p}}{(q-1)\sinh u(\cosh u)^{q-1}} \nonumber\\
& & - \frac{p(p-1)}{(q-1)(q-2)}\int\frac{u^{p-2}}{(\cosh u)^{q-2}} du + 
\frac{q-2}{q-1}\int\frac{u^{p}}{(\cosh u)^{q-2}} du.
\end{eqnarray}
As a simple integration by parts yields: 
\begin{equation}
\int \frac{u^{p}}{(\cosh u)^{2}} du = u^{p}\tanh u - p\int u^{p-1}\tanh
u\,du,
\end{equation}
we clearly see that to accomplish our calculation we have to find the
integral: 
\begin{equation}
\int u^{p}\tanh u\,du
\end{equation}
which requires special treatment. It will be proven below that it is
expressible in terms of the polylogarithm functions $\mathrm{Li}_{n}(z),
n=1, 2,..., p+1,$ with a properly chosen argument $z$. There is considerable
literature on analytical and numerical properties of these functions~\cite{Levin1,Levin2}
being a generalization of Euler's dilogarithm. In the unit circle
polylogarithm of integral order $m>1$ can be defined by 
\begin{equation}
\mathrm{Li}_{m}(z) = \sum_{k=1}^{\infty}\frac{z^{k}}{k^{m}}.
\end{equation}
From this series we easily derive the important recurrence relation: 
\begin{equation}
\frac{d\:}{dz}\mathrm{Li}_{n}(z) = \frac{1}{z}\mathrm{Li}_{n-1}(z)
                                                  \label{recc}
\end{equation}
which will also be used in the integral form: 
\begin{equation}
\mathrm{Li}_{n}(z) = \int\frac{\mathrm{Li}_{n-1}(z)}{z} dz +C.
\end{equation}
Using this property and the integral representation of dilogarithm 
\begin{equation}
\mathrm{Li}_{2}(z)=-\int_{0}^{z}\frac{\ln(1-t)}{t}\,dt
\end{equation}
one can extend the definition of polylogarithms to low orders: 
\begin{equation}
\mathrm{Li}_{1}(z)=-\ln(1-z),\:\:\mathrm{Li}_{0}(z)=\frac{z}{1-z}.
                                                               \label{AA}
\end{equation}
Starting from the recurrence relation in the integral form we repeatedly
integrate by parts using the expression for the derivative of a
polylogarithm (\ref{recc}) at every step: 
\begin{eqnarray}
\mathrm{Li}_{n}(z)& =& C+\int (\ln z)^{\prime}\,\mathrm{Li}_{n-1}(z) dz =...=\nonumber
\\
&=& C+\ln z\,\mathrm{Li}_{n-1}(z) -\frac{1}{2}(\ln z)^{2}\,\mathrm{Li}
_{n-2}(z) +\frac{1}{2\cdot 3}(\ln z)^{3}\,\mathrm{Li}_{n-3}(z)+... \nonumber \\
& & +\frac{(-1)^{n-1}}{(n-2)!}(\ln z)^{n-2}\,\mathrm{Li}_{2}(z) +\frac{
(-1)^{n-1}}{(n-1)!}(\ln z)^{n-1}\,\ln(1-z) +\frac{(-1)^{n-1}}{(n-1)!}\int
\frac{(\ln z)^{n-1}}{1-z}dz,\nonumber\\
                    \label{AAA}
\end{eqnarray}
where in the final step we have taken into account the form of $\mathrm{Li}
_{1}(z)$. If we now set $z=-e^{-2u}$ we recover the required integral in the
last term: 
\begin{equation}
(-1)^{n-1}\int\frac{(\ln z)^{n-1}}{1-z}dz = 
2^{n}\int u^{n-1}\frac{e^{-u}}{e^{u}+e^{-u}}du = 2^{n-1}
\left[\frac{u^{n}}{n} -\int u^{n-1}\tanh u\,du \right].
\end{equation}
After rearrangement of the sum in the identity (\ref{AAA}) and change $
n\!-\!1\rightarrow p$ we get the expression: 
\begin{eqnarray}
\int u^{p}\tanh u\,du &=& C + \frac{u^{p+1}}{p+1} + u^{p}\ln(1+e^{-2u}) -
\sum_{k=1}^{p}\frac{p!}{2^{k}(p-k)!}\,u^{p-k}\,\mathrm{Li}_{k+1}(-e^{-2u}).\nonumber\\
\end{eqnarray}
The hyperbolic functions and monomials in $u$ variable arising in our
calculation can also be cast in the unified way when the special forms of
low-order polylogarithms (\ref{AA}) are used.\newline
The case studied here falls into a wide range of expressions which can be
integrated with the help of polylogarithms. Some occurrences of these
functions in physical problems are mentioned in the Levin's book \cite
{Levin1}. It is remarkable that they arise in Feynman diagrams integrals, in
particular in computation of quantum electrodynamics corrections to the
electron gyromagnetic ratio \cite{Laporta}.\\[5mm]

%\bibliography{prd,abg}

\end{document}